\documentclass[aps,prl,twocolumn,superscriptaddress,amsmath,amssymb,longbibliography]{revtex4-2}
\usepackage[english]{babel}
\usepackage{graphicx,bbm}
\usepackage[utf8]{inputenc}
\usepackage[colorlinks,linkcolor=blue,citecolor=blue,urlcolor=blue]{hyperref}
\usepackage{bbold}
\usepackage{physics}
\usepackage[normalem]{ulem}

\newcommand{\bfp}{\mathbf{p}}
\newcommand{\bfh}{\mathbf{h}}

\begin{document} 
\title{Long-living prethermalization in nearly integrable spin ladders}

\author{J. Pawłowski}
\affiliation{Institute of Theoretical Physics, Faculty of Fundamental Problems of Technology, Wroc{\l}aw University of Science and Technology, 50-370 Wroc{\l}aw, Poland}

\author{M. Panfil}
\affiliation{Faculty of Physics, University of Warsaw, Pasteura 5, 02-093 Warsaw, Poland}

\author{J. Herbrych}
\affiliation{Institute of Theoretical Physics, Faculty of Fundamental Problems of Technology, Wroc{\l}aw University of Science and Technology, 50-370 Wroc{\l}aw, Poland}

\author{M. Mierzejewski}
\affiliation{Institute of Theoretical Physics, Faculty of Fundamental Problems of Technology, Wroc{\l}aw University of Science and Technology, 50-370 Wroc{\l}aw, Poland}

\date{\today}
\begin{abstract}
Relaxation rates in nearly integrable systems usually increase quadratically with the strength of the perturbation that breaks integrability. We show that the relaxation rates can be significantly smaller in systems that are integrable along two intersecting lines in the parameter space. In the vicinity of the intersection point, the relaxation rates of certain observables increase with the fourth power of the distance from this point, whereas for other observables one observes standard quadratic dependence on the perturbation. As a result, one obtains exceedingly long-living prethermalization but with a reduced number of the nearly conserved operators. We show also that such a scenario can be realized in spin ladders. 
\end{abstract}
\maketitle

{\it Introduction.} The time evolution of generic quantum systems tends towards the thermal equilibrium \cite{Trotzky2012,Kaufman2016,Neill2016, DAlessio2016, Rigol2008thermalization} independently of the initial state. In recent years, systems in which thermalization occurs very slowly \cite{tang_kao_18} or can be completely eliminated \cite{Kinoshita2006} have attracted a lot of interest. Particular attention was paid to integrable systems which avoid thermalization and evolve towards a generalized Gibbs state \cite{rigol2007,cassidy2011,lange2018, Ilievski2015, langen2015experimental}. The crucial role in the behavior of such systems is played by local (or quasilocal) integrals of motion (LIOMs) whose presence prevents thermalization of local observables \cite{mazur,zotos1997} and has important consequences for the transport properties of integrable systems \cite{bertini2021, Bertini2016, Ilievski2017, DeNardis2021}. However, more realistic models as well as experimental setups contain small, but non-negligible, perturbations which break the integrability \cite{prosen1998,zotos2004,jung06,jung07,huang2013,essler2014,Brandino2015}. While one expects that asymptotic dynamics of such nearly integrable (NI) systems is diffusive \cite{LeBlond2020,Znidaric2020,Bastianello2020}, the dynamics at intermediate time-scales resembles that of integrable models. The latter transient dynamics of NI systems is known as prethermalization \cite{kollar2011,bertini2015,mallayya2019,bastianello2021}. 

A particularly important example of the integrability breaking occurs in systems of weakly coupled integrable chains \cite{tang_kao_18, Caux2019}. While the interchain coupling can be well controlled in the cold-atom experiments \cite{Bordia2016} 
it is not always possible to completely eliminate this interaction \cite{kao2021}. Quite obviously, a nonvanishing interchain coupling is  unavoidable in solid-state systems \cite{scheie2021, gannon2019spinon}. Moreover, recent quasiclassical studies based on the Boltzmann collision integral approach \cite{Panfil2023,lebek2023} indicate that extremely long relaxation times may occur in such NI systems. 

It is rather obvious that one is most interested in NI systems in which the relaxation times are as long as possible. While an NI system may host very distinct relaxation times \cite{mierzejewski2015,friedman2020,bastianello2021,mierzejewski2022}, the corresponding relaxation rates typically scale quadratically with the strength of the integrability-breaking perturbation \cite{jung06,jung07,mierzejewski2015,mallayya2018,mierzejewski2022}. Under such a scenario, the only way to increase the relaxation times is to reduce the perturbation. In this Letter, we establish other possibility of decreasing the relaxation rates in NI systems. Namely, we consider a system that is integrable along two intersecting lines in the parameter space, see, e.g., Refs. \cite{DELFINO2006291,Kormos_Nat2017,fogarty2021} for an example of such systems. 
If certain LIOMs on both lines have large overlaps, then the corresponding relaxation rates increase with the fourth power of the distance (in the parameter space) from the intersection point. Relaxation rates for LIOMs that do not have such overlaps, exhibit standard quadratic dependence on the perturbation. As a consequence, extremely small relaxation rates and arbitrary larger ratios of relaxation times appear in the studied NI system. Finally, we show that such a scenario can be implemented in nearly integrable spin ladders introduced below.  

{\it Spin ladder.} 
We investigate a spin ladder consisting of two XXZ chains coupled via anisotropic spin-spin interaction of strength $U$
\begin{align}
 H &= \sum_{\ell=1}^2 H_{\ell} + U \sum_{j = 1}^{L} S^{z}_{j,1} S^z_{j,2}\; , \label{ham} \\
 H_{\ell} &= \frac{J}{2} \sum_{j= 1}^{L} \left( S^{+}_{j,\ell} S^{-}_{j+1,\ell} + {\rm H.c.} \right)\label{haml}
 + \Delta \sum_{j = 1}^{L} S^{z}_{j,\ell} S^{z}_{j+1,\ell }\;. 
\end{align}
The subscripts $\ell=1,2$ and $j=1,...,L$ denote, respectively, the leg and the site within a leg on which the \mbox{spin-$1/2$} operators act. From now on we set \(J=1\), fix the total magnetization to \(S^z_{\text{tot}} = 0\) and assume periodic boundary conditions along the legs of the ladder.

\begin{figure}[!htbp]
 \includegraphics[width=\columnwidth]{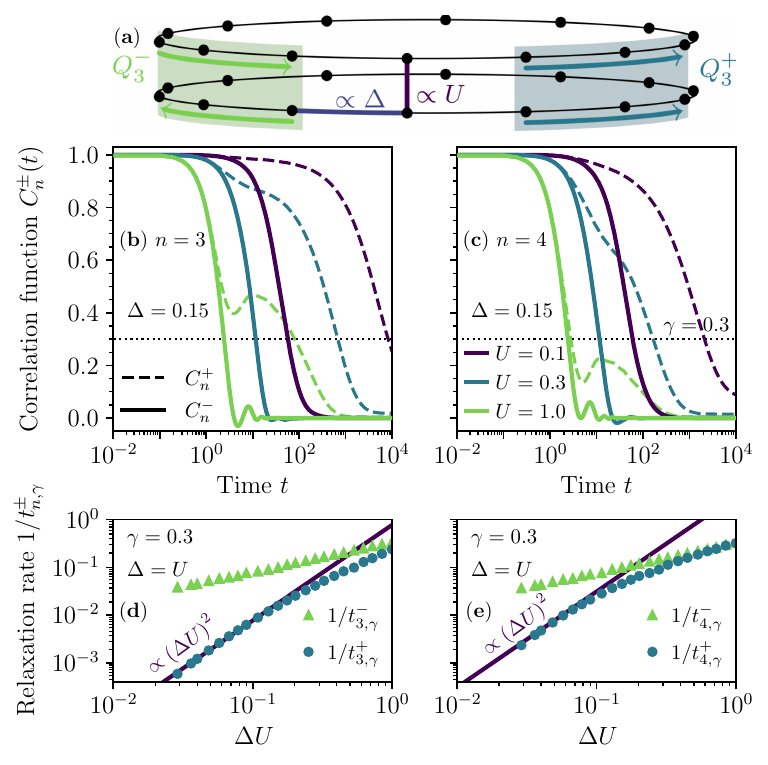}
 \caption{(a) Sketch of the ladder in Eq.~(\ref{ham}) with marked \mbox{$S^z$-$S^z$} interactions ($\Delta, U$) and nearly conserved operators ($Q^{\pm}_n$).(b,c) Correlation functions $C^{\pm}_n(t)$ defined in Eq.~(\ref{corfun}) obtained for a ladder with $L=14$ rungs. (b) Dashed and continuous curves show $C^+_3(t)$ and $C^-_3(t)$, respectively. (c) The same as in (b) but for $C^{\pm}_4(t)$. Horizontal dotted lines in (b,c) define times, $t^{\pm}_{n,\gamma}$, when $C^{\pm}_n(t^{\pm}_{n,\gamma})=\gamma=0.3$. Legends in (b,c) are common for both panels. Panels (d,e) show $1/t^{\pm}_{3,\gamma}$ and $1/t^{\pm}_{4,\gamma}$ for NI system with $U=\Delta$ and $\gamma=0.3$ (see text for details).}
 \label{fig:sum_diff}
\end{figure}

The ladder is shown schematically in Fig.~\ref{fig:sum_diff}(a). It is integrable for \(U = 0\) and inherits a complete set of LIOMs \(\{Q_{n,1}\}\) and \(\{Q_{n,2}\}\) from both XXZ-chains. { In this work we use analytic forms of LIOMs  and the notation from Ref~\cite{Grabowski1995}. In particular, }
\(Q_{1,\ell}\) and \(Q_{2,\ell}\) denote respectively the total magnetization and the Hamiltonians of the chain $\ell$. Here, we focus on the dynamics of the first two nontrivial XXZ LIOMs, namely \(Q_{3,\ell}\) and \(Q_{4,\ell} \), supported on \(3\) and \(4\) sites, respectively. This choice is motivated by that \(Q_{3,\ell}\) is the energy current and thus it is an experimentally relevant quantity. In order to demonstrate that the discussed properties are not unique to just a single quantity, we study also \(Q_{4,\ell} \). 

It is convenient to introduce symmetrized combinations of the latter LIOMs, \(Q^{\pm}_{n} \equiv Q_{n,1} \pm Q_{n,2}\). In the case of uncoupled chains, \(Q^{+}_{n} \) as well as \(Q^{-}_{n} \) are strictly conserved thus the correlation functions are time-independent, i.e. $\langle Q^{\pm}_{n}(t) Q^{\pm}_{n} \rangle=$ const, {  and we use a simplified notation $Q^{\pm}_{n}\equiv Q^{\pm}_{n}(t=0)$.} However, the interaction term, $U\ne 0$, breaks the integrability of the studied model so that $\langle Q^{\pm}_{n}(t) Q^{\pm}_{n} \rangle$ decay in time. In the following we show that the sums of the XXZ LIOMs, $Q^{+}_{n}$, decay much slower than their differences, $Q^{-}_{n}$. We present an explanation of this unexpected behavior, by inspecting a dual point of view, in which the intrachain term $\propto \Delta$ is also treated as an integrability-breaking perturbation. Namely for \(\Delta = 0\), the Hamiltonian of the ladder reduces to the Hubbard chain, in which the leg index $\ell$ labels the spin projection of fermions. This view introduces another set of LIOMs \(\{I_n\}\), originating from the integrability of the Hubbard chain. Here, we argue that the decay of \( Q_n^{+} \) or \( I_n \) in the NI model ($U \ne 0,\Delta \ne 0$) is significantly slowed down due large overlaps of both sets of LIOMs. 

{\it Dynamics of nearly conserved observables.}
To probe the dynamics of the nearly integrable spin ladder, we calculate the real-time correlation functions
\begin{equation}
C^{\pm}_n(t)= \langle e^{iHt} Q_n^{\pm} e^{-iHt} Q^{\pm}_n \rangle\;. \label{corfun}
\end{equation}
Here, \(\langle A B \rangle = \frac{1}{\mathcal{Z}} \mathrm{Tr}(AB)\) is the Hilbert-Schmidt inner product for Hermitian operators $A$, $B$ and \(\mathcal{Z} = \mathrm{Tr}(\mathbbm{1})\) is the dimension of the Hilbert space. We recall that the Hilbert-Schmidt product is mathematically equivalent to the ensemble average at infinite temperature. 

We note that \(Q_{4,\ell}\) and \(I_{4}\) in Ref~\cite{Grabowski1995} are not orthogonal to the respective integrable Hamiltonians, $H_{\ell}$ and $H(\Delta=0)$. Therefore, we first subtract their projections on the Hamiltonians and obtain orthogonal sets of LIOMs. All considered LIOMs are also Hilbert-Schmidt normalized, i.e. \(\lVert Q_n^{\pm} \rVert^2 = \langle Q_n^{\pm}Q_n^{\pm} \rangle = 1\), and thus the correlation functions in Eq. (\ref{corfun})  are equal to one at \(t=0\). We refer to Supplemental Material~\cite{supmat} for explicit forms of LIOMs and their overlaps.

Utilizing the Lanczos time evolution method \cite{Park1986,mierzejewski2010} combined with the dynamical typicality \cite{Bartsch2009,Gemmer2009,Elsayed2013,Steinigeweg2014a,Steinigeweg2015} we calculate correlation functions introduced in Eq.~(\ref{corfun}), see Ref.~\cite{supmat} for the details of numerical calculations. Figs.~\ref{fig:sum_diff}(b) and \ref{fig:sum_diff}(c) show, respectively,  $C^{\pm}_{3}(t)$ and $C^{\pm}_{4}(t)$ calculated for small anisotropy \(\Delta = 0.15,\) and different strengths of the interchain interaction, \(U = 0.1,0.3,1 \). In the regime of small $U$ one observes that the correlation functions obtained for $Q^+_n$ (dashed lines) decay much slower than the correlation functions determined for $Q^-_n$ (continuous lines). In the Supplemental Material \cite{supmat} we show that the differences between $C^+_n(t)$ and $C^-_n(t)$ become significant for much shorter times than the time-scale at which $C^+_n(t)$ develop the finite-size effects. Therefore, the exceedingly different relaxation times for $Q^{+}_n$ and $Q^{-}_n$ do not emerge as finite-size artifacts. 

In order to capture the differences between relaxation of $Q^+_n$ and $Q^-_n$ in a quantitative manner, we determine times when the correlation functions decay to a fraction of \(\gamma\) of their initial value, such that \mbox{\( C^{\pm}_n(t^{\pm}_{n,\gamma}) = \gamma \)}, see dotted lines in Fig.~\ref{fig:sum_diff}(b,c). While the accessible system sizes do not allow us to reliably establish the true relaxation rates, we assume that their dependence on $U$ and $\Delta$ can be estimated from $t^{\pm}_{n,\gamma}$. In Fig.~\ref{fig:sum_diff}(d,e) we show the corresponding relaxation rates $1/t^{\pm}_{3,{\gamma}}$ and $1/t^{\pm}_{4,{\gamma}}$ for an NI system along the line $U=\Delta$ where we set $\gamma=0.3$. In the regime of weak interactions, one observes that the relaxation rates for $Q^{+}_n$ increase only as $U^2 \Delta^2$, i.e., they are much smaller than the squared strengths of integrability-breaking interactions $U^2$ or $\Delta^2$. However, the relaxation rates for the other set of nearly conserved operators, $Q^{-}_n$, show much weaker dependence on perturbations and may be larger than $1/t^{+}_{n,\gamma}$ by a few orders of magnitude. 

\begin{figure}[!htbp]
 \centering
 \includegraphics[width=\columnwidth]{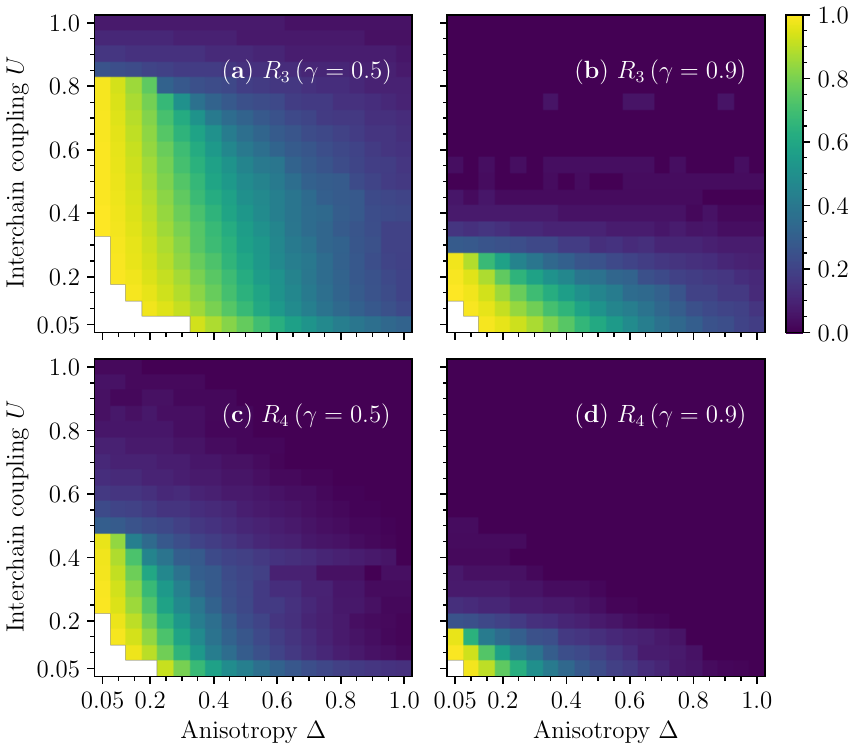}
 \caption{Ratio \(R_n(\gamma)\) calculated for \(L=12\). Blank regions corresponds to parameters for which $C^+_n(t) > \gamma$ for the numerically accessible times $t \sim 10^4$, e.g., see $C^+_n(t)$ for \(U=0.1\) in Fig.~\ref{fig:sum_diff}(b).}
 \label{fig:ttd}
\end{figure}

Next we check how the differences between $t^{+}_{n,\gamma}$ and $t^{-}_{n,\gamma}$ depend on the parameters of the studied model. To this end we calculate the ratio \mbox{\(R_n(\gamma) = (t^+_{n,\gamma} - t^-_{n,\gamma})/(t^+_{n,\gamma} + t^-_{n,\gamma})\)}. Numerical results for this ratio are shown in Fig.~\ref{fig:ttd} on an evenly-spaced rectangular grid in the parameter space \((\Delta,U)\). Blank parts on the plots correspond to the situation when $t^+_n$ is larger than the longest time accessible in our numerical calculations, $t\sim 10^4$. For small \( \Delta \) and \(U\) we observe that $R_n(\gamma)$ is close to one for both $n=3$ and $n=4$. It means that the decay times for $Q^-_n$ are negligibly small when compared to the time-scale that corresponds to the slow relaxation of $Q^+_n$. For large \( \Delta \) and \(U\) we observe that $Q^+_n$ and $Q^-_n$ relax rather quickly and with roughly the same relaxation times, $t^{+}_{n,\gamma} \simeq t^{-}_{n,\gamma}$.

{\it Significance of overlapping LIOMs.} In order to explain the origin of the exceedingly different and long relaxation times, we turn to a dual picture. Namely, we consider the anisotropy term ($\sim \Delta$) as a perturbation to the integrable Hubbard chain described by the Hamiltonian $H(\Delta = 0)$. The latter Hamiltonian possesses another complete set of LIOMs \(\{I_n\}\). In what follows, we demonstrate that the slower decay of the operators \( Q_n^{+} \) in the nearly integrable ladder ($\Delta \ne 0$ and $U\ne 0$) can be linked to their substantial overlaps with \(I_n\). Such overlaps do not exist for the quickly decaying operators, \( Q_n^{-} \). We note that \( Q_n^{-} \) are odd under the spin-flip transformation, $\ell \to 3-\ell$, whereas the Hubbard LIOMs, \(I_n\), are even under such spin-flip so that one obtains $\langle Q_n^{-} I_n \rangle=0 $.

\begin{figure}[!htb]
 \includegraphics[width=1.0\columnwidth]{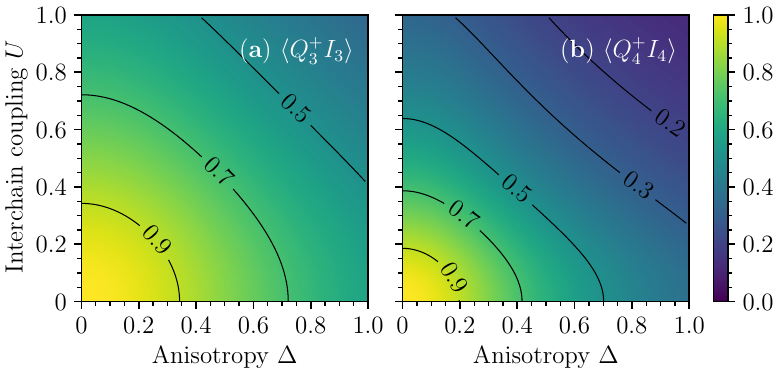}
 \caption{Overlaps $\langle Q_n^{+} I_n \rangle$ calculated analytically for $L \to \infty$, see Eq.~(\ref{eq:Q3_overlap}) and Ref. \cite{supmat} for more details. Black solid curves are isolines. Note that $Q_n^{+}$ and $I_n$ are strictly conserved for $U=0$ and $\Delta=0$, respectively.}
 \label{fig:overlaps}
\end{figure}

In Fig.~\ref{fig:overlaps} we present the overlaps $\langle Q_n^{+} I_n \rangle$. In order to completely eliminate the finite-size effects, the overlaps were calculated analytically in the full Hilbert space that includes all $S^z_{\rm tot}$-sector. In particular, one finds
\begin{equation}
 \langle Q_3^{+}I_3 \rangle = \frac{J^2}{\sqrt{\left(J^2 + 2U^2\right)\left( J^2+2\Delta^2 \right) }}\;,
 \label{eq:Q3_overlap}
\end{equation} 
and the explicit form of the other overlap $\langle Q_4^{+}I_4 \rangle$ is shown in Ref. \cite{supmat}. We have also checked that the numerically obtained overlaps in the sector with $S^z_{\rm tot}=0$  (not shown) are qualitatively the same as the results in Fig.~\ref{fig:overlaps}. Comparing Fig.~\ref{fig:ttd} with Fig.~\ref{fig:overlaps} we find that the differences in relaxations of $ Q_n^{+}$ and $ Q_n^{-}$ are most pronounced for the same parameters where the overlaps $\langle Q_n^{+} I_n \rangle$ are large.

Finally, we establish a simple link between the overlaps of LIOMs of integrable models ($U=0$ or $\Delta=0$) and the slow dynamics of $Q^{+}_n$ in the nearly integrable ladder with ($U\ne 0$ and $\Delta \ne 0$). To this end we conjecture that in the regime of small $U$ and $\Delta$, the relaxation rates for $Q^{+}_n$ and $Q^{-}_n$ can be expanded in powers of $\Delta^2$ and $U^2$. Since $Q^{-}_n$ are strictly conserved only for $U=0$ and arbitrary $\Delta$, the lowest-order contributions to their relaxation rates are $1/t^-_{n,\gamma} \propto U^2$, as it is expected for a generic integrability-breaking perturbation. However, due to large overlaps $\langle Q^{+}_{n} I_n \rangle$, the relaxation rates for $Q^{+}_{n}$ vanish both for \mbox{$U=0$} as well as for $\Delta=0$. Therefore these relaxation rates cannot contain terms which depend solely on either $U$ or $\Delta$ thus the lowest order contributions are \mbox{$1/t^+_{n,\gamma} \propto \Delta^2 U^2$}. Consequently, for small $\Delta$ one obtains $t^+_{n,\gamma} \gg t^-_{n,\gamma}$.

\begin{figure}[!htb]
 \centering
 \includegraphics[width=\columnwidth]{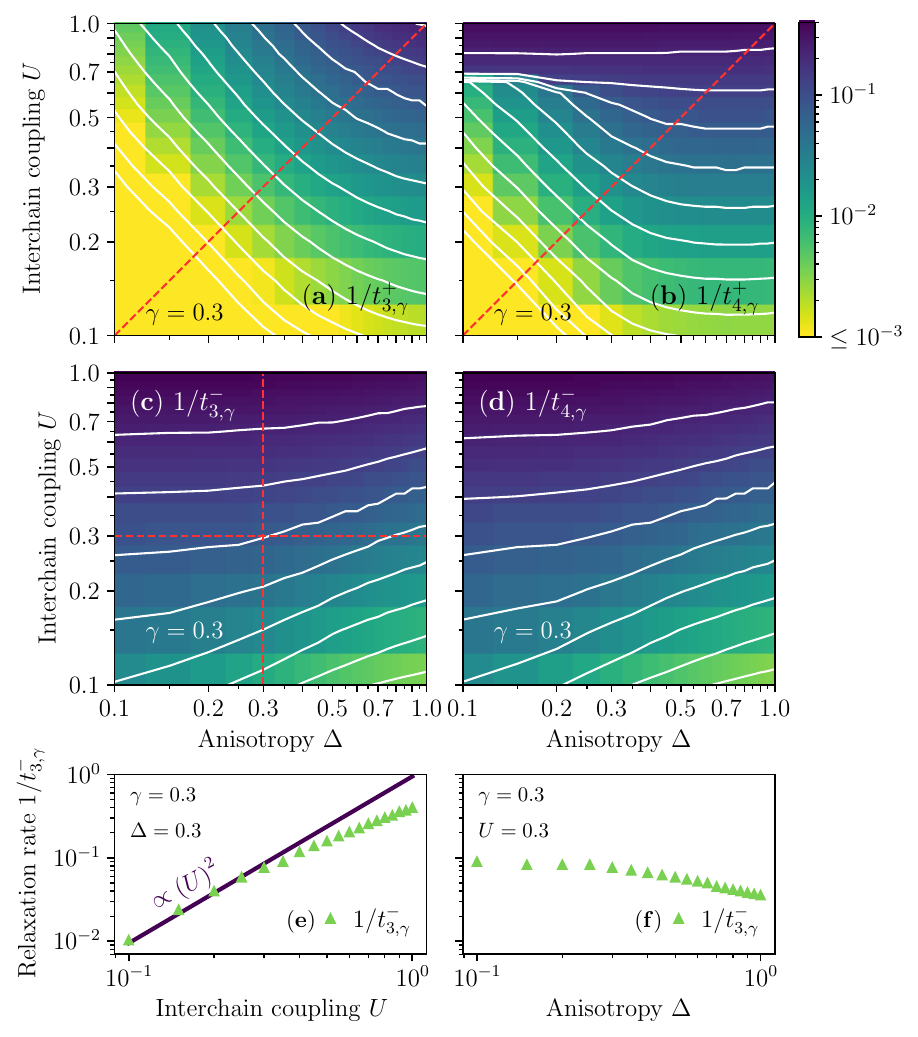}
 \caption{(a)-(d) Relaxation rates $1/t^{\pm}_{n,\gamma}$ estimated from the correlation functions in 
 Eq.~(\ref{corfun}) via relation
 $C^{\pm}_n(t^{\pm}_{n,\gamma})=\gamma $ for $\gamma=0.3$ and $L=12$. Continuous curves represent isolines. Dashed lines in (a) and (b) mark parameters for which relaxation rates are shown in Fig.~\ref{fig:sum_diff}(d,e). Panels (e) and (f) show relaxation rates $1/t^{-}_{n,\gamma}$ along the dashed lines marked in (c).}
 \label{fig:inv_tau}
\end{figure}

This scenario is clearly confirmed by results in Figs.~\ref{fig:inv_tau}(a,b). In these plots we show heat maps for $1/t^{+}_{n,\gamma}$ using logarithmic scales for $U$ and $\Delta$. One observes that the isolines roughly follow straight lines consistent with the dependence $1/t^{+}_{n,\gamma} \propto (\Delta U)^{\alpha} =$ const. Numerical results obtained in a direction that is perpendicular to the isolines ($U=\Delta$) are shown in Fig.~\ref{fig:sum_diff}(d,e) demonstrating that the exponent $\alpha=2$. Using parametrization $U=d\cos(\phi),\; \Delta=d\sin(\phi)$, we find that the relaxation rate for $Q^+_n$ grows as $d^4$. Here, $d$ is distance from the intersection of two lines, $\Delta=0$, and $U=0$, along which the studied model is integrable. 

In the case of $1/t^-_n$, one observes very different isolines with positive slopes, see Figs.~\ref{fig:inv_tau}(c,d). The latter are consistent with the conjecture that $1/t^-_{n,\gamma}$ are determined mostly by the interchain interaction $U$. For the sake of completeness we have calculated $1/t^-_{n,\gamma}$ in the directions that are roughly perpendicular or parallel to the corresponding isolines, i.e., for $\Delta=$ const or  $U=$ const. Numerical results shown in Fig.~\ref{fig:inv_tau}(e,f) confirm the standard quadratic dependence of $1/t^-_{n,\gamma}$ on the perturbation $U$.

Since the proximity of two integrable lines is responsible for the long-living prethermalization in the studied ladder, one may expect to find a broader class of operators which exhibit slow relaxation. In the Supplemental Material \cite{supmat} we show that linear combinations of $Q^+_n$ and $I_n$ show very similar dynamics. { As expected, breaking of integrability along one of the lines (e.g. via
other form of the coupling between the legs)  destroys the exceptional properties of the studied system \cite{supmat}}.

{\it Conclusions} We have considered a ladder consisting of two XXZ chains (each with spin anisotropy $\Delta$) coupled via interaction of strength $U$. The studied model is integrable along two lines in the parameter space. Namely for $\Delta=0$ the ladder represents the Hubbard chain with one set of LIOMs $\{I_n\}$, whereas for $U=0$ one obtains two uncoupled XXZ chains. In the latter case we have introduced LIOMs which are symmetric, $\{Q^+_n \}$, or antisymmetric, $\{Q^-_n\}$, with respect to exchanging the chains. Studying the dynamics of a nearly integrable ladder with ($U\ne 0$ and $\Delta \ne 0$) we have found that correlation functions for $Q^+_n$ decay much slower than for $Q^-_n$ and that the difference of relaxation times is most pronounced for small $U$ and $\Delta$. 

We have linked this result with large overlaps between $Q^+_n$ and $I_n$ and vanishing overlaps between $Q^-_n$ and $I_n$. As a consequences of the former overlaps, the relaxation rates for $Q^+_n$ must vanish for both $U=0$ and $\Delta=0$ so that the lowest-order contribution to the relaxation rates is at most of the order of $ \Delta^2 U^2$. In contrast to this, the relaxation rates for $Q^-_n$ are of the order of $U^2$. Such behaviour explains exceedingly different relaxation times observed for $Q^+_n$ and $Q^-_n$ in the regime of small $U$ and $\Delta$. Consequently, in this regime of parameters one deals with a rather specific prethermalization. Namely, the number of nearly conserved quantities, $Q^+_{n}$, is twice smaller than the number of LIOMs in the uncoupled chains, where both $Q^+_{n}$ and $Q^-_{n}$ are conserved.

These findings can be further examined from the point of view of quasiclassical analysis based on the Boltzmann collision integral{~\cite{PhysRevLett.127.130601,Panfil2023} (see also~\cite{friedman2020,bastianello2021} for the hydrodynamic perspective on the integrability breaking). The central assumption here is that each leg of the spin chain is in a state described by the Generalized Gibbs Ensemble (GGE) of the XXZ spin chain~\cite{PhysRevLett.113.187203}. The whole system is then in the product state of the two GGE states. The coupling between the legs leads then, by the Fermi's golden rule, to an evolution of states of each leg.} As we show in the Supplemental Material~\cite{supmat}, both families of charges $Q_n^{\pm}$ are conserved { during this evolution} up to order $\Delta^2 U^2$ in a case when { the anisotropy parameters and the states of the two legs are identical}. Otherwise, $Q_n^{-}$ may acquire dynamics at lower orders while $Q_n^{+}$ does not. Our results reinforce this quasiclassical picture.

Our reasoning is general and { is expected to hold true also for other systems which are integrable along two intersecting lines in the parameter space. However, this  should be verified by direct calculations.} The essential conditions is that the LIOMs on both integrable lines have large mutual overlaps in the vicinity of the crossing point. Then, in the vicinity of this point one may expect long-living prethermalization with relaxation rates that increase with the fourth power of the distance in the parameter space from the intersection point. This is in contrast to the case of generic nearly integrable systems where relaxation rates increase with second power of the perturbation.  { The long-living prethermalization  in the studied  ladder leads to  a very slow relaxation of the energy current \cite{supmat} that should also be visible  as a   high thermal conductivity.}

\begin{acknowledgments}
M.M. and J.P. acknowledge support by the National Science Centre (NCN), Poland via project 2020/37/B/ST3/00020. M.P. acknowledges support by the National Science Centre (NCN), Poland via project 2022/47/B/ST2/03334. Part of the calculations has been carried out using resources provided by Wroclaw Centre for Networking and Supercomputing (\url{http://wcss.pl}).
\end{acknowledgments}

\bibliographystyle{biblev1}
\bibliography{manuintegrals}

\clearpage
\appendix
\setcounter{page}{1}
\setcounter{figure}{0}
\setcounter{equation}{0}
\newcommand{\rom}[1]{\uppercase\expandafter{\romannumeral #1\relax}}
\renewcommand{\refname}{Supplemental References}
\renewcommand{\figurename}{Supplemental Figure}
\renewcommand{\thefigure}{S\arabic{figure}}
\renewcommand{\theHfigure}{S\arabic{figure}} 
\renewcommand{\citenumfont}[1]{#1}
\renewcommand{\bibnumfmt}[1]{[#1]}
\renewcommand{\thepage}{S\arabic{page}}
\renewcommand{\theequation}{S\arabic{equation}}
\renewcommand{\theHequation}{S\arabic{figure}} 
\renewcommand{\thetable}{S\Roman{equation}}
\renewcommand{\theHtable}{S\arabic{equation}} 
\onecolumngrid
\begin{center}
{\large \bf Supplemental Material:\\
Long-living prethermalization in nearly integrable spin ladders}\\

\vspace{0.3cm}

\setcounter{page}{1}

\ J. Pawłowski$^{1}$, M. Panfil$^{2}$, J. Herbrych$^{1}$, M. Mierzejewski$^{1}$
\\
\ \\
$^1${\it Institute of Theoretical Physics, Faculty of Fundamental Problems of Technology, \\ Wroc\l aw University of Science and Technology, 50-370 Wroc\l aw, Poland} \\
$^2${\it Faculty of Physics, University of Warsaw, Pasteura 5, 02-093 Warsaw, Poland}\\

\author{M. Panfil}
\affiliation{Faculty of Physics, University of Warsaw, Pasteura 5, 02-093 Warsaw, Poland}

\author{J. Herbrych}
\affiliation{Institute of Theoretical Physics, Faculty of Fundamental Problems of Technology, Wroc{\l}aw University of Science and Technology, 50-370 Wroc{\l}aw, Poland}

\end{center}

\vspace{0.6cm}
In the Supplemental Material we discuss overlaps of the local integrals of motion (LIOMs).
We also provide technical details of numerical calculations, discuss the finite-size effects and  dynamics of rotated integrals of motion. Finally, we analyze our numerical results from the perspective of the Boltzmann equation.\\
\vspace{0.3cm}

\twocolumngrid
\section{Overlaps of the local integrals of motion}
In the main text, we have shown analytical expression for the overlap between Heisenberg and Hubbard LIOMs (Eq.~\ref{eq:Q3_overlap}). Here, we provide additional details on how these formulas are obtained and derive analogous expression for $\langle Q^+_4 I_4 \rangle$. For completeness, let us first recall the form of LIOMs, as derived in Ref.~\cite{Grabowski1995},
\begin{align}
 \widetilde{Q}^{\prime}_{3,\ell} =& \,\frac{iJ}{2} \sum_{j=1}^{L} \bigg( J S_{j,\ell}^{-} S_{j+1,\ell}^{z} S_{j+2,\ell}^{+} + \Delta S_{j,\ell}^{z} S_{j+1,\ell}^{+} S_{j+2,\ell}^{-}\nonumber\\
 + & \Delta S_{j,\ell}^{+} S_{j+1,\ell}^{-} S_{j+2,\ell}^{z} \bigg) + \mathrm{H.c.},
\end{align}

\begin{align}
 \widetilde{I}^{\prime}_3 =& \, \frac{iJ}{4} \sum_{j=1}^{L} \bigg( J S_{j,1}^{-}S_{j+1,1}^{z}S_{j+2,1}^{+} + U S_{j,1}^{+}S_{j+1,1}^{-}S_{j,2}^{z} \nonumber \\
 + & U S_{j,1}^{+}S_{j+1,1}^{-}S_{j+1,2}^{z} + [1 \leftrightarrow 2] \bigg) + \mathrm{H.c.},
\end{align}

\begin{widetext}
\begin{align} 
 \widetilde{Q}^{\prime}_{4,\ell} = \frac{J}{4} \sum_{j=1}^{L} \bigg( & J^2 S_{j,\ell}^{+}S_{j+1,\ell}^{-} + 4J^2 S_{j,\ell}^{+}S_{j+1,\ell}^{z}S_{j+2,\ell}^{z}S_{j+3,\ell}^{-} 
 - J\Delta S_{j,\ell}^{+}S_{j+2,\ell}^{-} - 2 J\Delta S_{j,\ell}^{+}S_{j+1,\ell}^{-}S_{j+2,\ell}^{-}S_{j+3,\ell}^{+} \nonumber\\
 + & 2 J\Delta S_{j,\ell}^{+}S_{j+1,\ell}^{-}S_{j+2,\ell}^{+}S_{j+3,\ell}^{-} + 2J\Delta S_{j,\ell}^{z}S_{j+1,\ell}^{z}
 - J\Delta S_{j,\ell}^{z}S_{j+2,\ell}^{z} - 4J\Delta S_{j,\ell}^{z}S_{j+1,\ell}^{+}S_{j+2,\ell}^{z}S_{j+3,\ell}^{-} \nonumber\\
 - & 4J\Delta S_{j,\ell}^{+}S_{j+1,\ell}^{z}S_{j+2,\ell}^{-}S_{j+3,\ell}^{z} + \Delta^2 S_{j,\ell}^{+}S_{j+1,\ell}^{-}
 + 4\Delta^2 S_{j,\ell}^{z}S_{j+1,\ell}^{+}S_{j+2,\ell}^{-}S_{j+3,\ell}^{z} \bigg) + \mathrm{H.c.},
\end{align}
\begin{align}
 \widetilde{I}^{\prime}_4 = \frac{J}{4} \sum_{j=1}^{L} \bigg( & 4J^2 \big(S_{j,1}^{+}S_{j+1,1}^{z}S_{j+2,1}^{z}S_{j+3,1}^{-} + S_{j,1}^{-}S_{j+1,1}^{z}S_{j+2,1}^{z}S_{j+3,1}^{+}\big) 
 - 2J U S_{j,1}^{z}S_{j+1,2}^{z} - J U S_{j,1}^{z}S_{j,2}^{z} \nonumber \\
 + & JU\big( S_{j,1}^{+}S_{j+1,1}^{-} - S_{j,1}^{-}S_{j+1,1}^{+} \big) \big[ \big( S_{j,2}^{+}S_{j+1,2}^{-} - S_{j,2}^{-}S_{j+1,2}^{+} \big)
 + 2 \big( S_{j+1,2}^{+}S_{j+2,2}^{-} - S_{j+1,2}^{-}S_{j+2,2}^{+} \big) \big] \nonumber \\ 
 - &4 J U \big(S_{j,1}^{+}S_{j+1,1}^{z}S_{j+2,1}^{-} + S_{j,1}^{-}S_{j+1,1}^{z}S_{j+2,1}^{+} \big) \big( S_{j,2}^{z} + S_{j+1,2}^{z} + S_{j+2,2}^{z} \big) \nonumber \\
 + &U^2 \big(S_{j,1}^{+}S_{j+1,1}^{-} + S_{j,1}^{-}S_{j+1,1}^{+}\big) \big( 4S_{j,2}^{z}S_{j,2}^{z} + \mathbbm{1}_{j,2}\mathbbm{1}_{j+1,2} \big) 
 + [1 \leftrightarrow 2]\bigg),
\end{align}
\end{widetext}
where \([1\leftrightarrow 2]\) denotes all previous terms, but with swapped leg index. We use the tilde and prime symbols to mark LIOMs which are not normalized and not orthogonal, respectively. 

Heisenberg LIOMs are generalized to the ladder setting as in the main text, \(\widetilde{Q}_n^{\prime,\pm} = \widetilde{Q}^{\prime}_{n,1} \pm \widetilde{Q}^{\prime}_{n,2}\). We also introduce Hilbert-Schmidt normalized LIOMs, \(I^{\prime}_n = \widetilde{I}^{\prime}_n/||\widetilde{I}^{\prime}_n||\) and \(Q^{\prime,\pm}_n = \widetilde{Q}^{\prime,\pm}_n/||\widetilde{Q}^{\prime,\pm}_n||\). The evaluation of the overlap of current-like LIOMs is straightforward, as they are already orthogonal to all lower-order LIOMs, $I_2,I_1$ or $Q^{\pm}_2, Q^{\pm}_1$. Consequently, $ I_3=I^{\prime}_3$ and $Q_3^{\pm}=Q^{\prime,\pm}_3$ and the overlap
\begin{equation}
 \langle Q_3^{+} I_3 \rangle = \frac{1}{\mathcal{Z}} \mathrm{Tr} \left(Q_3^{+} I_3 \right) = \frac{1}{\mathcal{Z}}\frac{\mathrm{Tr}\left(\widetilde{Q}_3^{+}\widetilde{O}_3\right)}{ ||\widetilde{Q}_3^{+}|| ||\widetilde{I}_3||},
 \label{eq:olap_details}
\end{equation}
yields Eq.~\eqref{eq:Q3_overlap} in the main text. To calculate the trace in the above expression, we collect all terms from the product of operators and use the fact that the spin operators are traceless. We also note that the trace over the full Hilbert space (\(\mathcal{Z}=4^L\)) factorizes over  sites and legs of the ladder. Finally, values of non-vanishing terms such as \(\mathrm{Tr}(S_{j,\ell}^{z}S_{j,\ell}^z)\), are obtained using the properties of the Pauli matrices.

Case of the \(\langle Q_4^{+} I_4 \rangle\) overlap is more complicated, as the charges \(\widetilde{Q}^{\prime,+}_4\) and \(\widetilde{I}^{\prime}_4\) must first be orthogonalized with respect to the lower-order LIOMs, i.e. the Hamiltonians \(H(U=0)\) and \(H(\Delta = 0)\) respectively. We do this in the usual fashion, by subtracting the projections
\begin{align}
 \widetilde{Q}_4^{+} & = \widetilde{Q}_4^{\prime,+} - \frac{\langle \widetilde{Q}_4^{\prime,+} H(U=0) \rangle}{\langle H(U=0) H(U=0) \rangle} H(U=0) \nonumber \\
 & = \widetilde{Q}_4^{\prime,+} - \frac{4\left(J^4+2J^2\Delta^2\right)}{2J^2+\Delta^2} H(U=0), \label{eq:Q4_proj} \\
 \widetilde{I}_4 & = \widetilde{I}_4^{\prime} - \frac{\langle \widetilde{I}_4^{\prime} H(\Delta=0) \rangle}{\langle H(\Delta=0) H(\Delta=0) \rangle} H(\Delta=0) \nonumber \\
 & = \widetilde{O}_4^{\prime} - \frac{3JU^2}{8J^2+2U^2} H(\Delta=0). \label{eq:O4_proj}
\end{align}
Carrying out the same calculation (including normalization of operators) as for Eq.~\eqref{eq:olap_details}, but with LIOMs given by~\eqref{eq:Q4_proj} and~\eqref{eq:O4_proj} one arrives at the expression 
\begin{widetext}
 \begin{equation}
 \langle Q_4^{+}I_4 \rangle = \frac{2 \left(8 J^8+2 J^6 \left(2 \Delta ^2+U^2\right)+\Delta ^2 J^2 U^2 \left(\Delta ^2-U^2\right)+\Delta ^4 U^4\right)}{\sqrt{\left(8 J^8+56 \Delta ^2 J^6+48 \Delta ^4 J^4+19 \Delta ^6 J^2+4 \Delta ^8\right) \left(4 J^2+U^2\right) \left(8 J^6+56 J^4 U^2+25 J^2 U^4+4 U^6\right)}}\;.
 \label{eq:Q4_overlap}
 \end{equation}
\end{widetext}

\section{Numerical methods}
We evaluate the correlation functions, defined in Eq.~\eqref{corfun} in the main text, using Quantum Typicality \cite{Bartsch2009,Gemmer2009}. Such approach has been successfully applied to various quantum many-body systems, in particular to the spin chains~\cite{Elsayed2013,Steinigeweg2014a,Steinigeweg2015}. The essence of this approach is to approximate the trace over the full Hilbert space with an expectation value in a random pure state drawn from a suitable ensemble
\begin{equation}
 C_n^{\pm}(t) \simeq \matrixel{\psi}{e^{iHt} Q_n^{\pm} e^{-iHt} Q^{\pm}_n}{\psi}\;.
 \label{typicality}
\end{equation}
Imposing unitary invariance on the distribution of the random states \(\ket{\psi} = \sum_{j=1}^{\mathcal{Z}} c_j\ket{j}\) and assuming independence of all \(\Re c_j\) as well as \(\Im c_j\), yields a Gaussian distribution for the coefficients, \(\Re c_j\) and \(\Im c_j\), in arbitrary orthonormal basis \(\{\ket{j}\}\)~\cite{Gemmer2009}. 
We then have two crucial properties~\cite{Bartsch2009,Steinigeweg2015} for the mean value and the variance of results obtained for various realizations of $\{ c_i \}$:

\begin{align}
 & \mathbb{E}_{\ket{\psi}}\left[\matrixel{\psi}{Q_n^{\pm}(t)Q^{\pm}_n}{\psi}\right] = C^{\pm}_n(t)\\
 & \sigma\left[\matrixel{\psi}{Q_n^{\pm}(t) Q^{\pm}_n}{\psi}\right] \leq
 {\cal O} \left(\frac{1}{\sqrt{\mathcal{Z}}} \right)
\end{align}

Hence, contribution of a single pure state can already be an exponentially good approximation of \(C_n^{\pm}(t)\), which for small systems can be further improved by additional sampling. We have checked that such improvement does not lead to noticeable changes of results shown in the present work and for the studied problems one may use only a single random state for each calculation. In our studies we shifted the time evolution to two auxiliary pure states \(\ket{\psi}\), \(\ket{\phi^{\pm}_n}=Q^{\pm}_n\ket{\psi}\) and calculated it using the Lanczos time evolution~\cite{Park1986} with a time step \(\delta t \leq 1\) and \(M=20\) Lanczos steps.

\begin{figure}[!htbp]
 \includegraphics[width=\columnwidth]{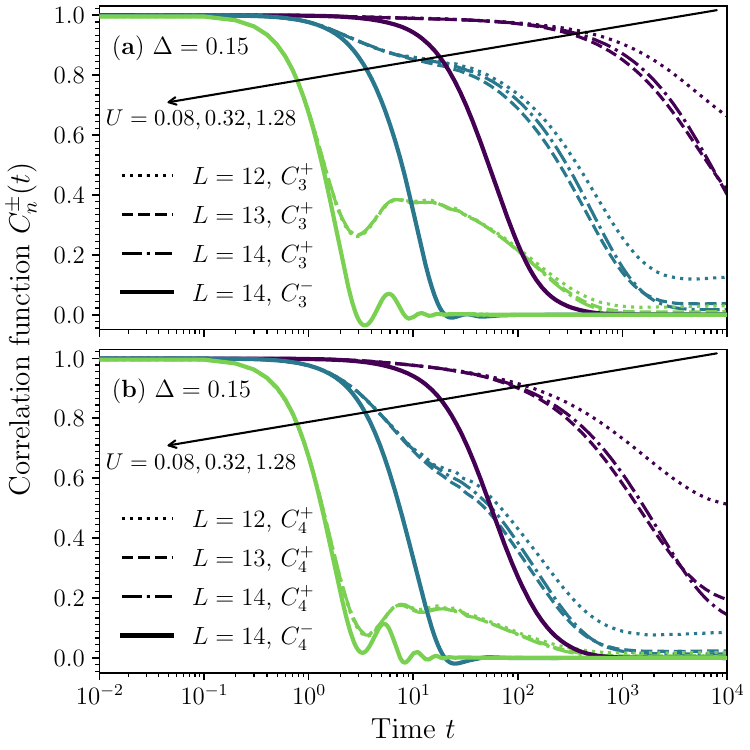}
 \caption{Correlation functions defined in Eq.~(\ref{corfun}) in the main text. Panel (a) depicts $C^+_3(t)$ for $L=12,13,14$ compared to $C^-_3(t)$ for $L=14$, while panel (b) the same as (a) but for $C^{\pm}_4(t)$. Arrows mark the values of \mbox{$U=0.08,0.32,1.28$} in ascending order.}
 \label{fig:fss}
\end{figure}

\section{Finite-size effects}
Since the ladder geometry strongly restricts accessible numbers of rungs, it is important to show that our major results do not arise as finite-size artifacts. To this end in Fig.~\ref{fig:fss} we show the correlation functions defined in Eq.~(\ref{corfun}) in the main text. Results for $C^+_n(t)$ are shown for various $L$ whereas $C^-_n(t)$ is shown only for the largest $L$. We observe that significant differences between \(C^+_n(t)\) and \(C^-_n(t)\) are visible on timescales which are much shorter than the appearance of any serious finite-size effects for \(C^+_n(t)\).

\begin{figure}[!h]
 \centering
 \includegraphics[width=\columnwidth]{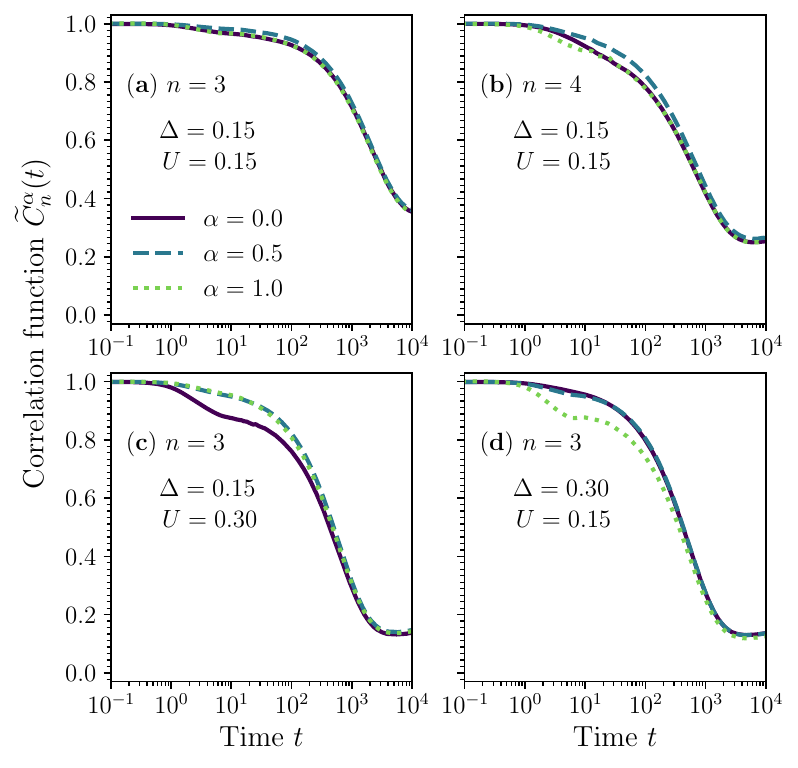}
 \caption{ Correlation function obtained for rotated operators defined in Eq.~(\ref{rot}), \mbox{$\widetilde{C}^{\alpha}_n(t)= \langle e^{iHt} O_n(\alpha) e^{-iHt} O_n(\alpha) \rangle$}. Results are obtained for $L=12$. Note that $\widetilde{C}^{0}_{n}=C^{+}_n(t)$ and $C^{+}_n(t)$ is 
 discussed in the main text.
 }
 \label{fig:mixing}
\end{figure}

\section{Rotation of the nearly conserved operators}
We have argued in the main text that the proximity of two integrable lines $U=0$ and $\Delta=0$ is responsible for long-living prethermalization that was obtained from slow relaxation of $Q^{+}_n$. However, such choice of slowly relaxing operators (from one out of two integrable lines) is arbitrary. Within our scenario, one expects similar dynamics also for $I_n$ or, more generally, for arbitrary combination of $Q^{+}_n$ and $I_n$. As a consistency check we confirm this expectation studying linear combinations of both types of LIOMs
\begin{equation}
 O_n(\alpha) = \frac{\cos(\alpha \frac{\pi}{2}) Q_n^{+} + \sin(\alpha \frac{\pi}{2}) I_n}{||\cos(\alpha\frac{\pi}{2}) Q_n^{+} + \sin(\alpha\frac{\pi}{2}) I_n ||} \; . \label{rot}
\end{equation} 
The parameter \(\alpha \in \left[ 0, 1 \right] \) controls the rotation between LIOMs of the Hubbard chain and those of the uncoupled XXZ chains. Fig.~\ref{fig:mixing} shows correlation functions $\widetilde{C}^{\alpha}_{n}(t)$ defined as in Eq.~(\ref{corfun}) in the main text but for $O_n(\alpha)$ instead of $Q^{\pm}_n$. Fig. \ref{fig:mixing}(a,b) show results obtained for equal parameters $U=\Delta$ when the dynamics of $O_n(\alpha)$ turns out to be insensitive to the rotation angle. Otherwise we observe that the larger parameter out of $\{U,\Delta\}$ sets the optimal observable, whereas the smaller one sets the integrablity breaking perturbation. In particular, for $\Delta < U$ in Fig.~\ref{fig:mixing}(c) and $\Delta > U$ in Fig.~\ref{fig:mixing}(d) we observe the slowest decay of $I_3$ and $Q^{+}_3$, respectively. However, the general conclusion for small $U$ and $\Delta$ is that the correlation function is roughly independent of the rotation introduced in 
Eq.~(\ref{rot}). 

\section{ The energy current}

\begin{figure}[!h]
 \centering
 \includegraphics[width=\columnwidth]{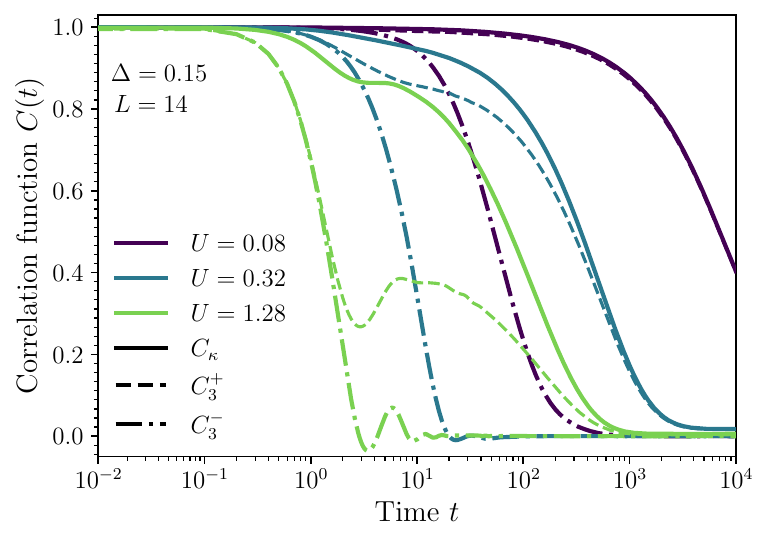}
 \caption{  Correlation function obtained for the energy current defined in Eq.~(\ref{corfunje}) along with  results for  $C^{\pm}_{3}(t)$, see Eq. (\ref{corfun}) in the main text.   Colors mark the value of the interchain interaction, $U$, and various types of lines correspond to different  correlation functions.  Results are obtained for $L=14$ and $\Delta=0.15$.
 }
 \label{fig:energy}
\end{figure}

{
The main conclusion following from our investigations concerns the long-living prethermalization. It originates from the integrability of the studied system along two intersecting lines  and  from large overlaps of LIOMs on these lines, $\langle Q^+_n I_n \rangle \sim 1$. Since $Q^+_3$ represents the  energy current of the unperturbed system ($U=0$) one may expect a nearly ballistic energy transport, or equivalently, high thermal conductivity for $U\ne 0$.  In order to directly confirm the former expectation we first calculate the energy current for the model defined in Eq. (\ref{ham}) in the main text. To this end we define the energy density \begin{eqnarray}
 h_j &=& \sum_{\ell=1}^2 \left[\frac{J}{2}\left(S^{+}_{j,\ell} S^{-}_{j+1,\ell} + {\rm H.c.} \right) +  \Delta S^{z}_{j,\ell} S^{z}_{j+1,\ell } \right] \nonumber \\ 
&&+ \frac{U}{2}\left(S^{z}_{j,1} S^z_{j,2}+ S^{z}_{j+1,1} S^z_{j+1,2}\right)\; ,
\end{eqnarray}
and introduce the corresponding polarization operator $P=\sum_j j h_j $. The energy current can be obtained from the time-dependence of P, namely $I_{\kappa}=i[H,P]$.   Direct calculations yield three contributions to the energy current $I_{\kappa}=I_{JJ}+I_{J\Delta}+I_{JU}$, where
\begin{eqnarray}
I_{JJ}&=&\frac{iJ^2}{2}\sum_{j=1}^{L}\left( S^{-}_{j,1} S^{z}_{j+1,1} S^{+}_{j+2,1} + [1 \leftrightarrow 2] \right)+ {\rm H.c.}\; ,\nonumber \\
&& \\
I_{J\Delta}&=& \frac{iJ \Delta }{2}\sum_{j=1}^{L} \left( S^{z}_{j,1} S^{+}_{j+1,1} S^{-}_{j+2,1} \right. \nonumber \\
&& \left.+ S^{+}_{j,1} S^{-}_{j+1,1} S^{z}_{j+2,1} +  [1 \leftrightarrow 2] \right)+ {\rm H.c.} \; , \\
I_{JU}&=& \frac{iJ U }{4}\sum_{j=1}^{L} \left( S^{+}_{j,1} S^{-}_{j+1,1} S^{z}_{j+1,2} \right. \nonumber \\
&& \left.+ S^{+}_{j,1} S^{-}_{j+1,1} S^{z}_{j,2} +  [1 \leftrightarrow 2] \right)+ {\rm H.c.} \;, 
 \end{eqnarray} 
and $  [1 \leftrightarrow 2] $ denotes the swap of the leg-indexes (i.e., the second indexes) in the preceding terms.
Figure \ref{fig:energy} shows numerical results for the normalized correlation function
\begin{equation}
C_{\kappa}(t)= \frac{\langle e^{iHt} I_{\kappa} e^{-iHt} I_{\kappa} \rangle}{\langle I_{\kappa}  I_{\kappa} \rangle}\;, \label{corfunje}
\end{equation}
along with the previously discussed  results for $C^{\pm}_{3}(t)$, see Eq. (\ref{corfun}) in the main text for the definition of the latter quantities.
One observes that the energy current inherits slow dynamics from $Q^+_3$. In particular for larger $U$ the dynamics of $I_{\kappa}$ is even slower than
that of $Q^+_3$. The latter result most probably originates from the (large?) overlap of the energy current with $I_3$.
}

\section{ Other form of the integrability-breaking perturbation}
{
The discussed mechanism of the long-living prethermalization is not expected to be robust against other forms of the interchain couplings. A generic coupling may either break the integrability for $\Delta=0$ or
it may reduce the overlaps of LIOMs on two integrable lines. In order to explicitly demonstrate such case, we have calculated the correlation functions  $C^{\pm}_n(t)$ for a modified Hamiltonian, cf. Eq. (\ref{ham}) in the main text, 
\begin{equation}
 H' = \sum_{\ell=1}^2 H_{\ell} + J_{\perp} \sum_{j = 1}^{L}  \frac{1}{2}\left( S^{+}_{j,1} S^-_{j,2} + {\rm H.c.} \right)\;. \label{hammod}
 \end{equation}
 Figure \ref{fig:energymod} shows the same results as figures  \ref{fig:sum_diff}(b) and  \ref{fig:sum_diff}(c) in the main text but for the interchain coupling introduced in Eq. (\ref{hammod}). 
 In contrast to the previously discussed model, we do not observe any substantial differences between the dynamics of $Q^+_{n}$ and $Q^-_{n}$ as shown in panels (a) and (b) for
 $n=3$ and $n=4$, respectively. Both quantities reveal similar relaxation times. These relaxation times are much smaller then  the relaxation times of  $Q^+_{n}$ obtained for the Hamiltonian  (\ref{ham}) from the main text.
 
\begin{figure}[!h]
 \centering
 \includegraphics[width=\columnwidth]{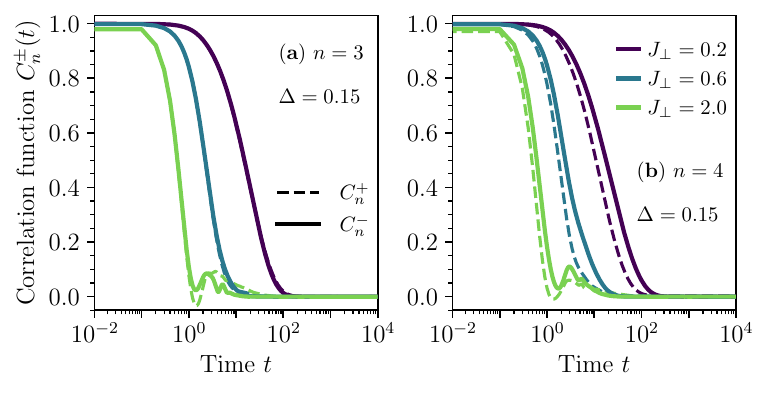}
 \caption{  The same results as in figures  \ref{fig:sum_diff}(b) and  \ref{fig:sum_diff}(c) in the main text but for the Hamiltonian (\ref{hammod}).  Results are obtained for $L=14$ and $\Delta=0.15$.
 }
 \label{fig:energymod}
\end{figure}
}

\section{Boltzmann equation}
We present here analysis of the dynamics of the two-leg XXZ spin chain based on the Boltzmann equation. We show that this quasiclassical analysis is consistent with the full quantum dynamics presented in the main text. Namely, the Boltzmann equation predicts that the dynamics occurs at order $\Delta^2 U^2$ or higher. Distinctively from the quantum case, both charges $Q^{\pm}_{n}$ are conserved with the same precision $\Delta^2 U^2$. It is then due to quantum effects that $Q^-_{n}$ attains contributions of order $U^2$, while $Q^+_{n}$ remains conserved with the precision predicted by the Boltzmann equation.

The Boltzmann approach to weakly perturbed integrable systems was developed recently in~\cite{friedman2020,bastianello2021,PhysRevLett.127.130601,Panfil2023}. In this approach it is assumed that the state of the system, throughout the whole time, is a product state of states of the two legs. The state of each leg is then described by the Generalized Gibbs Ensemble (GGE). The basic ingredient of the GGE for the XXZ spin chain is a set of density functions $\{\rho_{\rm p, a}(\theta)\}$. They describe the density of quasiparticles of type $a$ present in the state as a function of the quasimomentum $\theta$ commonly referred to as rapidity. Rapidity parametrises momentum $k_a(\theta)$ and energy $\omega_a(\theta)$ of a quasiparticle. Along the densities of particles it is common to introduce the densities of holes $\{\rho_{\rm h, a}(\theta)\}$. All those functions are determined from the densities of particles through the generalized thermodynamic Bethe ansatz  equations~\cite{friedman2020}.

The excitations in the XXZ spin chain are organized into {\em strings} and we refer to~\cite{TakahashiBOOK} for more detailed discussion on them. The Boltzmann approach leads then to evolution equations for $\{\rho_{\rm p, a}(\theta)\}$ in each leg induced due to scattering events caused the integrability breaking perturbation. When states in both legs are different, this leads to a pair of coupled equations. However, if they are the same in the initial state, they remain identical for all times. In this case the problem reduces to a single equation,
\begin{align} \label{supp:Boltzmann}
 \partial\rho_{\rm p}^{(a)}(\theta) &= \mathcal{I}^{(a)}[\rho_{\rm p}](\theta) = \sum_b Q^{(a,b)}[\rho_{\rm p}](\theta),
\end{align}
which is a straightforward generalization of the Boltzmann equation for two coupled Lieb-Liniger models~\cite{PhysRevLett.127.130601,Panfil2023} to a system hosting multiple types of excitations.

The collision integral $\mathcal{I}^{(a,b)}(\theta)$ describes two effects. First, following from the Fermi's golden rule, it takes into account all the processes that due to the perturbation modify the density of excitations of type $a$ at rapidity $\theta$ while creating excitation of type $b$ in the other leg. This defines a bare collision integral $Q_0^{(a,b)}(\theta)$. Furthermore, due to the interactions present in the XXZ spin chains, a modification to density at $\theta$ modifies the densities also elsewhere. This effect is captured by a back-flow function and we refer to~\cite{PhysRevLett.113.187203} for its precise definition in the GGE context. Importantly, the back-flow vanishes for $\Delta = 0$. This implies that its effect is subleading and can be neglected as our aim here is to estimate the order of the leading contribution. The leading order is then determined fully by the bare collision integral $\mathcal{I}_0^{(a,b)}[\rho_{\rm p}]$.

As stated above, the bare collision integral follows from the Fermi's golden rule. It involves the (norm squared of) matrix elements of the perturbing operator $U \sum_{j=1}^L S_{j,1}^z S_{j,2}^z$. The state of the system is the product state of states in both legs and the computation of the matrix element reduces then to computation of form-factors of $S^z_{j}$. The $S^z_j$ operator conserves the total magnetization of the state and therefore its form-factors are non-zero only between states with the same magnetization. The possible excitations are then organized into magnetization-conserving particle-hole excitations (in the Bethe Ansatz description of the spin chain). The perturbation theory in $\Delta$ shows that the leading processes are single particle-hole excitations. We denote the corresponding form-factors $F_a(p,h) = \langle \rho_{\rm p}| S_0^z | \rho_{\rm p}, (h \rightarrow p)_a\rangle$ with $(p,h)$ being the rapidities of particle and hole respectively. The energy and momentum of the excited state is $\omega_a(p,h) = \omega_a(p) - \omega_a(h)$ and $k_a(p,h) = k_a(p) - k_a(h)$ respectively. This allows us to write down the leading contribution to the collision integral as 
\begin{widetext}
\begin{equation}
 \mathcal{I}_0^{(a,b)}[\rho_{\rm p}] = \int {\rm d}\bfp {\rm d}\bfh\, A^2(k) |F_a(p,h)|^2 \left(S_b^{zz}(-k, -\omega) - S_b^{zz}(k, \omega)\frac{\rho_{{\rm h}, a}(h) \rho_{{\rm p}, a}(p)}{\rho_{{\rm h}, a}(p) \rho_{{\rm p}, a}(h)} \right)\left( 1 + \mathcal{O}(U^2 \Delta^2)\right),
\end{equation}
\end{widetext}
where $A$ is the Fourier transform of the potential coupling the two spin chains and $S_b^{zz}(k, \omega)$ is the contribution to the dynamic structure factor from excitations of type $b$ such that the whole dynamic structure factor is $S^{zz}(k, \omega) = \sum_b S_b^{zz}(k, \omega)$. Such factorization of the dynamic structure factor is a straightforward consequence of the spectral representation of any two point function. Finally, the integration measure includes the densities of particles and holes, ${\rm d}\bfp {\rm d}\bfh = {\rm d}p \rho_{\rm h, a}(p) {\rm d}h \rho_{\rm p, a}(h)$. 

To estimate the leading order of the bare collision integral we consider the spectral representation of the dynamic structure factor. It again involves form-factors of $S_j^z$ operator and, by the same argument as above, in the leading order in $\Delta$ we need to consider only single particle-hole processes. A contribution from excitations of type $b$ is then
\begin{equation} 
 S_b^{zz}(k, \omega)\! = \!\!\int\! {\rm d}\bfp {\rm d}\bfh\, |F_b(p,h)|^2 \delta(k - k_b(p,h)) \delta(\omega - \omega_b(p, h)).
\end{equation}
Under the particle-hole symmetry of the form factor, each contribution $S_b^{zz}(k, \omega)$ obeys a relation similar to the detailed balance,
\begin{equation} \label{supp:detailed_balance}
 S_b^{zz}(-k,-\omega) = \frac{\rho_{{\rm h}, b}(\bar{p}) \rho_{{\rm p}, b}(\bar{h}) }{\rho_{{\rm h}, b}(\bar{h}) \rho_{{\rm p}, b}(\bar{p}) } S_b^{zz}(k,\omega),
\end{equation}
where $(\bar{p}, \bar{h})$ is determined from the kinematic constraint $k = k_b(\bar{p},\bar{h}), \omega = \omega_b(\bar{p},\bar{h})$.

With this result at our disposal, the collision integral is
\begin{widetext}
\begin{equation}
 \mathcal{I}_0^{(a, b)}[\rho_{\rm p}] = \int {\rm d}\bfp {\rm d}\bfh\, A^2(k) |F_a(p,h)|^2 S_b^{zz}(-k, -\omega)\left(1 - \frac{\rho_{{\rm h}, a}(h) \rho_{{\rm p}, a}(p)}{\rho_{{\rm h}, a}(p) \rho_{{\rm p}, a}(h)} \frac{\rho_{{\rm h}, b}(\bar{h}) \rho_{{\rm p}, b}(\bar{p}) }{\rho_{{\rm h}, b}(\bar{p}) \rho_{{\rm p}, b}(\bar{h}) } \right)\left(1 + \mathcal{O}(U^2 \Delta^2)\right),
\end{equation}
\end{widetext}
with $(\bar{p}, \bar{h})$ determined from the energy-momentum constraint $k_a(p,h) + k_b(\bar{p}, \bar{h}) = 0$ and $\omega_a(p,h) + \omega_b(\bar{p},\bar{h}) = 0$. This is the final expression for the single-particle hole contribution to the bare collision integral. As argued above this contribution is the leading one. We will show that this contribution is of order $U^2 \Delta^2$. The coupling term $A(k)^2$ is of order $U^2$ and therefore it remains to show that the this expression is also of order $\Delta^2$. 

We analyze first the case $a = b$. The only solution to the energy-momentum constraint is then $\bar{p} = h$, $\bar{h} = p$, in consequence, the collision integral vanishes identically. For the remaining cases of $a\neq b$ we can use a perturbative argument. For $\Delta=0$ the spectrum of the theory is that of the free fermions. This implies that the only excitations, in the XXZ language, are $1$-strings. Higher strings excitations are bound states of $1$-strings and appear due to the effective attractive interaction induced by $\Delta \neq 0$.  Therefore, changing the type of an excitation is a process for which the form-factor is at least of the order $\Delta$. This implies that the whole expression is at least of the order $\Delta^2$.

This analysis shows that semi-classically the evolution of the whole system occurs at the order $\Delta^2 U^2$. This implies that both combinations $Q_{n}^{\pm}$ are conserved. We note that the situation changes when the initial states of both spin chains are different, a setup studied recently for coupled Lieb-Liniger models~\cite{lebek2023}. In such situation, already at the level of the Boltzmann equation, the odd combinations $Q_{n}^-$ acquire dynamics, while $Q_{n}^+$ remain conserved with the precision $U^2/c^2$, where $U$ is again the strength of the coupling between two legs and $1/c$ is a small parameter controlling the deviation from the free fermionic point (thus playing the role of $\Delta$). 

\end{document}